\documentclass[prc]{revtex4}
\usepackage{graphicx}
\begin{document}
\title{Origins of Bulk Viscosity in Relativistic Heavy Ion Collisions}
\author{Kerstin Paech and Scott Pratt}
\affiliation{Department of Physics and Astronomy,
Michigan State University\\
East Lansing, Michigan 48824-1321}
\date{\today}

 
\begin{abstract}
A variety of physical phenomena can lead to viscous effects. Several sources of shear and bulk viscosity are reviewed with an emphasis on the bulk viscosity associated with chiral restoration and with chemical non-equilibrium. We show that in a mean-field treatment of the limiting case of a second order phase transition, the bulk viscosity peaks in a singularity at the critical point.
\end{abstract}

\maketitle

\section{Introduction and Theory}
\label{sec:intro}
Viscosity has attracted remarkable attention in the RHIC (Relativistic Heavy Ion Collider) community during the first years of running. In particular, experimental observations of large elliptic flow have pointed to a small shear viscosity and inspired the term ``perfect liquid'' \cite{RHIC_Whitepapers}. In this paper, we review the general theoretical definition of viscosity, then show how five different physical effects can lead to non-zero viscous coefficients (Sec. 2). We focus on bulk viscosity, and show that one can find large, even singular, effects in the neighborhood of $T_c$ (Sec. 3). Although the present study focuses on understanding the behavior and physical explanation of the coefficients, we speculate on the experimental manifestations large viscosities might bring about.

In non-viscous hydrodynamics the elements of the stress-energy tensor depend only on the energy density $\epsilon$ and particle-number densities $\vec{n}$,
\begin{equation}
\widetilde{T}_{ij}^{\rm (non.visc.)}({\bf r},t)
= P(\epsilon,\vec{n})\delta_{ij},
\end{equation}
where $P$ is the pressure and $\epsilon$ and $\vec{n}$ are implicitly functions of ${\bf r}$ and $t$. The tilde denotes that $T_{ij}$ is evaluated in a frame where the collective velocity ${\bf u}({\bf r})=0$. In Navier-Stokes hydrodynamics, viscosity is incorporated by altering $\widetilde{T}$ so that it includes terms proportional to the velocity gradients $\partial u_i/\partial r_j$.
\begin{equation}
\label{eq:NS}
\tilde{T}_{ij}^{\rm (N.S.)} = P(\epsilon,\vec{n})\delta_{ij}
+\eta(\epsilon,\vec{n})\left(\frac{\partial u_i}{\partial r_j}+
\frac{\partial u_j}{\partial r_i}-\frac{2}{3}(\nabla\cdot{\bf u})\delta_{ij}\right)
+B(\epsilon,\vec{n})(\nabla\cdot{\bf u})\delta_{ij},
\end{equation}
Here, $\eta$ and $B$ are the shear and bulk viscosities. In ideal hydrodyanmics $\nabla\cdot {\bf u}=(\partial\epsilon/\partial t)/(P+\epsilon)$, so the bulk viscosity can be interpreted as describing the correction to the pressure linearly proportional to the rate at which the energy density changes, whereas the shear viscosity describes the asymmetry of $\widetilde{T}_{ij}$ due to an anisotropic expansion. In non-viscous hydrodynamics, accelerations are proportional to the gradient of the pressure, while in general, accelerations arise from derivatives of the stress energy tensor,
\begin{equation}
(\epsilon\delta_{ij}+\widetilde{T}_{ij})\frac{\partial u_j}{\partial t}=
-\frac{\partial}{\partial x_j}\widetilde{T}_{ij}.
\end{equation}
Thus, the components of the stress-energy tensor can be considered as representatives of the pressure in a given direction, and any reduction/rise of $\widetilde{T}_{ij}$ from viscous effects will result in a slowing/acceleration of the expansion in that direction.

Viscous coefficients can be expressed in terms of correlations in the stress-energy tensor through Kubo relations. These are derived by considering alterations of $T_{ii}$ due to a perturbation $V$. In linear response theory,
\begin{equation}
\delta\langle T_{ij}(r=0)\rangle
=-(i/\hbar)\int_{r'_0<0} d^4r' 
\langle[T_{ij}(r=0),V(r')]\rangle,~~~
V(r')=r'_i(\partial_iu_j) T_{0j}(r'),
\end{equation}
where the perturbation represents the change to the Hamiltonian due to boosting according to a linear velocity gradient.

To derive the Kubo relations, one first makes the substitution $T_{ij}\rightarrow \Delta T_{ij}$, where $\Delta T_{ij}$ refers to the difference with respect to the time-averaged value of $T_{ij}$. After inserting $(\partial_{r'_0} r'_0)$ into the integral one can integrate by parts, and because of the substitution with $\Delta T_{ij}$, one can dispose of the contributions at $r'_0=-\infty$ to obtain,
\begin{eqnarray}
\delta\langle T_{ij}(r=0)\rangle
&=&-(i/\hbar)\int_{r'_0<0} d^4r' (\partial_{r'_0}r'_0)
\langle[\Delta T_{ij}(r=0),\Delta T_{0j}(r')]\rangle r'_i(\partial_iu_j),\\
\nonumber
&=&
(i/\hbar)\int_{r'_0<0} d^4r' r'_0
\langle[\Delta T_{ij}(r=0),\partial_{r'_0}\Delta T_{0j}(r')]\rangle r'_i(\partial_iu_j),\\
\label{eq:deltij}
&=&(i/\hbar)\int_{r'_0<0} d^4r' r'_0
\langle [\Delta T_{ij}(r=0),\Delta T_{kl}(r')]\rangle
\partial_k u_l.
\end{eqnarray}
The last step involved applying the conservation of the stress-energy tensor, $\partial_tT_{i0}=-\partial_jT_{ij}$.

For $i\ne j$, symmetries constrain $k$ and $l$ to equal $i$ and $j$, which allows the extraction of the shear viscosity from Eq. (\ref{eq:NS}),
\begin{eqnarray}
\label{eq:kuboshear}
\eta&=&(i/\hbar)\int_{r'_0<0} d^4r' r'_0
\langle[\Delta T_{ij}(0),\Delta T_{ij}(r')]\rangle,~~~i\ne j\\
\nonumber
&=&\lim_{\omega\rightarrow 0} \frac{-1}{2\omega\hbar}\int d^4r' e^{i\omega t'}
\langle[\Delta T_{ij}(0),\Delta T_{ij}(r')]\rangle.
\end{eqnarray}
By considering the case where $\partial_iu_j=(1/3)\delta_{ij} \nabla\cdot{\bf u}$, one can inspect $T_{ii}$ in Eq. (\ref{eq:deltij}) to find the bulk viscosity,
\begin{eqnarray}
\label{eq:kuboB}
B&=&(i/3\hbar)\sum_j \int_{r'_0<0} d^4r' r'_0
\langle [\Delta T_{ii}(0),\Delta T_{jj}(r')]\rangle\\
\nonumber
&=&\lim_{\omega\rightarrow 0} \frac{-1}{6\omega\hbar}\sum_j
\int d^4r' e^{i\omega t'}
\langle [\Delta T_{ii}(0),\Delta T_{jj}(r')]\rangle.
\end{eqnarray}
The Kubo relations, Eq.s (\ref{eq:kuboshear}) and (\ref{eq:kuboB}), are fully consistent with quantum mechanics. The classical limit can be obtained by first noting that $\langle\cdots\rangle$ refers to a thermal average with temperature  $T=1/\beta$, then applying the identity \cite{foerster},
\begin{equation}
e^{-\beta H}V(t)=e^{i\beta\hbar\partial_t}V(t)e^{-\beta H},
\end{equation}
to one of the terms in the commutator in Eq.s (\ref{eq:kuboshear}) or (\ref{eq:kuboB}), then keeping the lowest term in $\hbar$,
\begin{equation}
{\rm Tr}~ e^{-\beta H}[\Delta T_{ij}(0),\Delta T_{kl}(r)]
\approx -i\hbar\beta{\rm Tr}~\partial_t \Delta T_{ij}(0)\Delta T_{kl}(r),\\
\end{equation}
which after an integration by parts gives the classical limit of the Kubo relations,
\begin{eqnarray}
\eta&\approx&\beta\int_{r'_0<0} d^4r'
\langle\Delta T_{ij}(0)\Delta T_{ij}(r')\rangle,~~~i\ne j\\
B&\approx&(\beta/3)\sum_j \int_{r'_0<0} d^4r'
\langle \Delta T_{ii}(0)\Delta T_{jj}(r')\rangle
\end{eqnarray}
The classical limit has been applied to determine the shear viscosity from simulations of molecular dynamics \cite{Muronga:2003tb,Gelman:2006xw}.

Although the Kubo relations are difficult to interpret physically, they do make it clear that viscosity is related to the size and to the damping of fluctuations of the elements $T_{ij}$. If fluctuations in $T_{ij}$ (at fixed energy) are large, or if they are slow to relax, a large viscosity will ensue.

Finally, since the Kubo relations are based on linear-response theory, i.e., assuming that the perturbation $V({\bf r})$ is small, we emphasize that the corrections to the stress-energy tensor represent an expansion in the velocity gradient. If the stress-energy tensor is strongly altered, it brings into question the validity of the linear approximation. This would be especially true if the viscous coefficients diverge as will be the case for the example discussed in Sec. \ref{sec:sigmabulk}.

\section{Five Sources of Viscosity}

Viscous effects arise whenever the elements of the stress-energy tensor, $T_{ij}$, have difficulty maintaining the equilibrium values due to a dynamically changing system, i.e., one with velocity gradients. In this section we briefly review five physical sources of viscosity, the first three of which have already been explained in the literature. Although viscosity in non-perturbative systems with ambiguous degrees of freedom might defy the simple descriptions of the five effects enumerated below, these examples provide physical insight into the richness of possible root causes for shear and bulk viscosity. 

\begin{enumerate}
\item {\bf Viscosity from non-zero mean free paths}: This is the most commonly understood source of viscosity. It is straight-forward to see how a non-zero collision time leads to an anisotropy for $T_{ij}$ by considering a velocity gradient for a Bjorken expansion, $u_z=z/\tau$, or equivalently, the velocity gradients are $\partial_zu_z=1/\tau, \partial_xu_x=\partial_yu_y=0$. We consider a particle whose momentum is $p'_z(\tau)$ when measured in the frame moving with the collective velocity corresponding to its position. In the absence of collisions, $p'_z$ will fall with $\tau$ since the particle will asymptotically approach a region where its velocity equals the collective velocity, $p'_z(\tau+\delta\tau)=p'_z(\tau)\tau/(\tau+\delta\tau)$. Meanwhile, $p'_x$ and $p'_y$ are frozen. The resulting anisotropy in the stress-energy tensor yields the following expression for the shear viscosity \cite{weinberg,Gyulassy:1997ib},
\begin{equation}
\eta=(4/5)P\tau_c,
\end{equation}
where $\tau_c$ is the collision time. The anisotropy increases the transverse pressure, giving radial flow an initial boost \cite{Teaney:2003kp}, and decreases the longitudinal pressure, thus reducing the longitudinal work which results in a larger transverse energy \cite{Gyulassy:1997ib}. It is also easy to see how such an expansion does not yield a bulk viscosity for either ultra-relativistic or non-relativistic gases. In those cases an isotropic expansion scales all three momenta proportional to $1/\tau$ which maintains thermal equilibrium, and collisions do not play a role. This is not the case when $m\sim T$, or especially if the gas has a mixture of relativistic and non-relativistic particles.

\item {\bf Viscosity from non-zero interaction range}: If the range of interaction between two-particles extends a distance $R$, interactions will share energy between particles from regions with different collective energies. A particle at $r=0$, where the collective energy is zero, will share energy with particles whose collective energy is $(1/2)m(R\partial_r u)^2$. For Boltzmann calculations, the viscosity will be proportional to $P R^2/\tau_c$ \cite{Cheng:2001dz}, with the constant of proportionality depending on the form of scattering kernel. Both bulk and shear terms result from non-zero interaction range. In Boltzmann calculations, the range of the interaction can approach zero for fixed scattering rates if the over-sampling ratio is allowed to approach infinity. Although this solves causality problems \cite{Kortemeyer:1995di}, it simultaneously eliminates viscous terms arising from finite-range scattering kernels, which might or might not be desirable. This has profound effects on calculations of elliptic flow, which can vary by a factor of 2 depending on the range of the scattering kernel \cite{Cheng:2001dz}.

\item {\bf Classical Electric Fields}: Color flux tubes form after the exchange of soft gluons between nucleons passing at high energy, and might also be formed during rapid hadronization. Additionally, longitudinal color electric fields might be created during the pre-thermalized stage of the collision (color-glass condensate). Since these fields tend to align with the velocity gradient, they can be a natural source of shear viscosity. In fact, if the fields are purely longitudinal, the elements of the stress-energy tensor become $T_{zz}=-\epsilon, T_{xx}=T_{yy}=\epsilon$. Thus, the transverse pressure becomes three times as stiff as a massless gas, $P=\epsilon/3$, which is usually considered a stiff equation of state. The negative longitudinal pressure signifies that the energy within a given unit of rapidity is increasing as the work done by the expansion is negative, similar to the stretching of a rubber band. This {\it hyper-shear} can lead to the development of early collective radial flow. A sophisticated  calculation including the effects of interactions amongst the fields \cite{Krasnitz:2002mn} showed a somewhat dampened anisotropy compared to the simple limit discussed here, with $T_{xx}=T_{yy}\approx 0.5\epsilon$ and $T_{zz}\approx 0$.

\item {\bf Non-equilibrium chemistry}: Chemical equilibirum can not be maintained unless the rate at which equilibrium abundances change is much smaller than the chemical equilibration rate, $1/\tau_{\rm chem}$,
\begin{equation}
\frac{dN}{dt}=-(1/\tau_{\rm chem})(N-N_{\rm eq}).
\end{equation}
If the equilbrium number is slowly changing, abundances will vary from equilbrium by an amount,
\begin{equation}
\label{eq:deltaN}
\delta N=-\tau_{\rm chem}\frac{dN_{\rm eq}}{dt}.
\end{equation}
To associate this departure from equilibrium as a viscosity, one must consider the corresponding change in the pressure,
\begin{equation}
\delta P=\left.\frac{\partial P}{\partial n}\right|_{{\rm fixed~}\epsilon}
\frac{\delta N}{\Omega},
\end{equation}
and make a connection between $dN_{\rm eq}/dt$ in Eq. (\ref{eq:deltaN}) with $\nabla\cdot{\bf u}$,
\begin{equation}
\frac{dN_{\rm eq}}{dt}=-\Omega\frac{\partial n}{\partial s}s\nabla\cdot{\bf u}.
\end{equation}
Here, $\Omega$ is the volume and the second relation exploits the fact that entropy is conserved in a slow expansion. The bulk viscosity is then found by comparison of the resulting change in pressure with the definition of viscosity in Eq. (\ref{eq:NS}),
\begin{eqnarray}
\delta P&=&\left.\frac{\partial P}{\partial n}\right|_{{\rm fixed~}\epsilon}
\frac{\partial n}{\partial s}s\tau_{\rm chem} (\nabla\cdot{\bf u}),\\
\nonumber
B&=&\left.\frac{\partial P}{\partial n}\right|_{{\rm fixed~}\epsilon}
\frac{\partial n}{\partial s}s\tau_{\rm chem}.
\end{eqnarray}
The bulk viscosity will be large whenever the equilibrium number is rapidly changing, e.g., the temperatures are falling below the masses, or masses are rising due to restoring chiral symmetry. If the hydrodynamic equations explicitly treat particle numbers as current obeying chemical evolution rates, chemical non-equilibration would not need to be accounted for through viscous terms.

\item {\bf Viscosity from dynamic mean fields}:
Bosonic mean fields, such as the $\sigma$ field, obey the Klein-Gordon equation. For fluctuations of wave number $k\rightarrow 0$,
\begin{eqnarray}
\label{eq:kg}
\frac{\partial^2}{\partial t^2}\Delta\sigma(t)
&=&-m_\sigma(T)^2\Delta\sigma(t)-\Gamma\frac{\partial}{\partial t}
\Delta\sigma(t),\\
\nonumber
\Delta\sigma(t)&\equiv&\sigma(t)-\sigma_{\rm eq}(\epsilon),
\end{eqnarray}
where $\sigma_{\rm eq}(\epsilon)$ is the equilibrium value of the condensate which is non-zero for $k=0$. The value of $\sigma_{\rm eq}$ is determined by minimizing the free energy, while the mass is related to the curvature of the free energy near the minimum,
\begin{equation}
\frac{\partial}{\partial \sigma}F(\sigma,T)=0,~~~
m_{\sigma}^2(T)
=\frac{\partial^2}{\partial \sigma^2}F(\sigma,T).
\end{equation}
One can see the equivalence of Eq. (\ref{eq:kg}) with the differential equation for the harmonic oscillator after performing the following substitutions,
\begin{equation}
k_{\rm h.o.}/m_{\rm h.o.}\rightarrow m^2_{\sigma},~~~~
\gamma_{\rm h.o.}/m_{\rm h.o.} \rightarrow \Gamma,
\end{equation}
where $\gamma_{\rm h.o.}$ is the drag coefficient for the harmonic oscillator, $k_{\rm h.o.}$ is the  spring constant and $m_{\rm h.o.}$ is the particle mass. The equivalence with the harmonic oscillator is shown in Appendix \ref{appendix} along with a derivation of the same result from the perspective of linear response theory assuming a Langevin force added to the equations of motion.

For the harmonic oscillator, the mean value of the position $x$ is altered if the equilibrium position is moving. The amount of the change was consistent with the drag force $\gamma dx_{\rm eq}/dt$ being equal and opposite to the restoring force $k\delta x$. The corresponding result can be derived for the damped Klein-Gordon equation,
\begin{equation}
\delta x=-\frac{\gamma_{\rm h.o.}}{k_{\rm h.o.}}\frac{dx_{\rm eq}}{dt},~~~~
\delta \sigma=-\frac{\Gamma}{m^2_{\sigma}(T)}\frac{d\sigma_{\rm eq}}{dt},
\end{equation}
where $\delta\sigma$ is the mean offset from the equilibrium value. Thus, $m^2_\sigma$ determines the restoring force, while $\Gamma$ describes the drag. Finite-size effects could be estimated by replacing $m^2$ with $m^2+k^2$, where $k^2$ would be given by the finite size, $k\sim 1/L$.  

The resulting bulk viscosity is,
\begin{equation}
B=\left.\frac{\partial P}{\partial \sigma}\right|_{{\rm fixed~}\epsilon}
\frac{\Gamma}{m_\sigma^2}\frac{\partial \sigma_{\rm eq}}{\partial s}s.
\end{equation}
The bulk viscosity is then large for energy densities where $\sigma$ is rapidly varying, or for when $m_\sigma$ is small, i.e., the critical region. 

\end{enumerate}

\section{Bulk Viscosity in the Linear Sigma Model}
\label{sec:sigmabulk}

Both of the last two sources of viscosity described in the previous section can be of special importance during the chiral transition. First, since masses are changing suddenly near $T_c$, chemical abundances should easily stray from equilibrium. Secondly, the mean field, which is zero above $T_c$ suddenly changes, and given the small masses in this region, large bulk viscosities are expected. 

\begin{figure}
\centerline{\includegraphics[width=8cm]{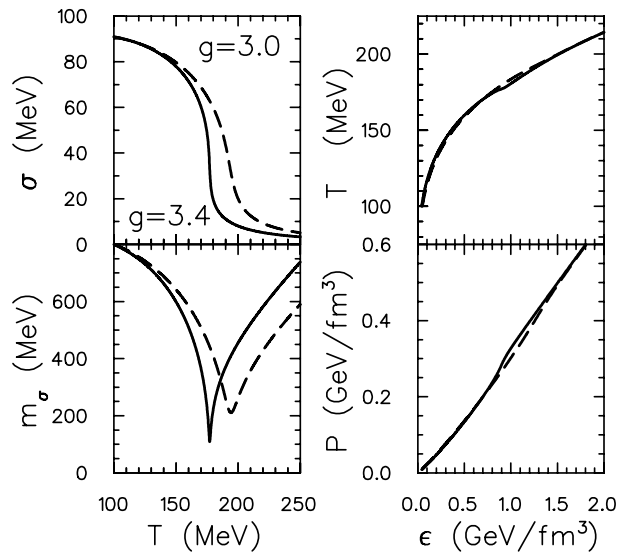}}
\caption{\label{fig:all4}For the linear sigma model, the sigma field and mass are shown as a function of the temperature in the left panels. Near $T_c$, the masses fall to zero and the mean value of the field changes rapidly, which gives rise to a sharp peak in the bulk viscosity. The pressure and temperature are displayed in the right-side panels as a function of energy density. The transition is sharper for $g=3.4$ (solid lines) which is near to the critical value, than for $g=3.0$ (dashed lines) which results in a smooth cross-over.
}
\end{figure}
As an example, we consider a simple example of a linear sigma model, where the coupling of the sigma field to the quarks provides the quark mass \cite{Paech:2005cx,Paech:2003fe},
\begin{equation}
H=-\frac{1}{2}\sigma\nabla^2\sigma+\frac{\lambda^4}{4}\left(
\sigma^2-f_\pi^2+m_\pi^2/\lambda^2\right)^2-h_q\sigma
+H_{\rm quarks}(m=g\sigma),
\end{equation}
assuming only up and down flavored quarks. The resulting equation of state and values for $m_\sigma$ and $\sigma$ are displayed in Fig. \ref{fig:all4} for $\lambda^2=40$. For couplings $g<g_c=3.554$, the transition is a smooth cross-over, while for $g=g_c$ the transition is second order, and for $g>g_c$ a first-order phase transition ensues with $T_c=172$ MeV. The values for $g_c$ and $T_c$ differ from \cite{Paech:2005cx} in that we neglect Fermi statistics here and also use a different value for $\lambda^2$. From Fig. \ref{fig:all4}, one can see that $m_\sigma$ becomes small in the same region that the field rapidly changes, which results in a peak in the bulk viscosity. For $g=g_c$, $m_\sigma^2\sim |T-T_c|$, and the viscosity behaves $\sim 1/|T-T_c|$. Since this is a mean field calculation we expect the critical exponent characterizing the singularity to differ from a more realistic quantum calculation.

To  calculate the bulk viscosity, we must evaluate the expressions from the previous section for $\delta n$ and $\delta \sigma$,
\begin{eqnarray}
\label{eq:deltansigma}
\delta n&=&-\tau_{\rm chem}\frac{dn_{\rm eq}}{ds}
s\nabla\cdot{\bf u},\\
\nonumber
\delta\sigma&=&-\frac{\Gamma}{m_\sigma^2}
\frac{d\sigma_{\rm eq}}{ds}s\nabla\cdot{\bf u}.
\end{eqnarray}
These translated into changes of the pressure, which can be calculated as a function of $\sigma$, $T$ and the chemical potential $\mu$, 
\begin{equation}
\label{eq:deltaP}
\delta P=\frac{\partial P}{\partial \mu}\delta\mu
+\frac{\partial P}{\partial T}\delta T
+\frac{\partial P}{\partial\sigma}\delta\sigma,
\end{equation}
where the last term is zero since $\sigma$ is chosen to minimize the free energy. The change in the pressure can then be found by solving  for $\delta\mu$ and $\delta T$ using knowledge of $\delta n$ from Eq. (\ref{eq:deltansigma}) and the condition that $\epsilon$ is fixed,
\begin{eqnarray}
\label{eq:deltane}
\delta \epsilon&=&\frac{\partial \epsilon}{\partial \mu}\delta\mu
+\frac{\partial \epsilon}{\partial T}\delta T
+\frac{\partial \epsilon}{\partial\sigma}\delta\sigma=0,\\
\nonumber
\delta n&=&\frac{\partial n}{\partial \mu}\delta\mu
+\frac{\partial n}{\partial T}\delta T
+\frac{\partial n}{\partial\sigma}\delta\sigma.
\end{eqnarray}
Solving for $\delta\mu$ and $\delta T$,
\begin{eqnarray}
\label{eq:bigmess}
\delta\mu&=&\frac{(\partial\epsilon/\partial T)\delta n +[(\partial\epsilon/\partial \sigma)(\partial n/\partial T)-(\partial\epsilon/\partial T)(\partial n/\partial\sigma)]\delta\sigma}
{(\partial\epsilon/\partial T)(\partial n/\partial\mu)-(\partial\epsilon/\partial\mu)(\partial n/\partial T)},\\
\nonumber
\delta T&=&\frac{(\partial\epsilon/\partial\mu)\delta n +[(\partial\epsilon/\partial\sigma)(\partial n/\partial\mu)-(\partial\epsilon/\partial\mu)(\partial n/\partial\sigma)]\delta\sigma}
{(\partial\epsilon/\partial\mu)(\partial n/\partial T)-(\partial\epsilon/\partial T)(\partial n/\partial\mu)}
\end{eqnarray}
The partition function and pressure for the quarks can be calculated analytically for fixed $\sigma$,
\begin{equation}
P_{\rm quarks}(m=g\sigma,T)=\frac{T\ln Z}{V}=
\frac{24}{2\pi^2}\left\{m^2T^2K_0(m/T)+2mT^3K_1(m/T))\right\},
\end{equation}
where the factor of 24 is the number of degrees of freedom. Analytic expressions for the derivatives in Eq. (\ref{eq:bigmess}) can also be found since derivatives of Bessel functions yield Bessel functions. 

Since equilibrium values are functions of $T$, the derivatives with respect to the entropy density in Eq.s (\ref{eq:deltansigma}) can be expressed in terms of $T$,
\begin{eqnarray}
\label{eq:equilderivs}
\frac{d\sigma_{\rm eq}}{ds}&=&
\frac{d\sigma_{\rm eq}/dT}
{ds/dT},\\
\nonumber
\frac{dn_{\rm eq}}{ds}&=&
\frac{dn_{\rm eq}/dT}
{ds/dT}.
\end{eqnarray}
The derivatives of $n_{\rm eq}$ and $\sigma_{\rm eq}$ in Eq.s (\ref{eq:equilderivs}) must be found numerically using the constraint that $\partial P/\partial\sigma=0$ at fixed $T$. After substituting the expressions for $d\sigma_{\rm eq}/dT$ and $dn_{\rm eq}/dT$ into Eq.s (\ref{eq:bigmess}), then substituting the resulting expressions for $\delta\mu$ and $\delta T$ into Eq. (\ref{eq:deltaP}), the resulting expression for $\delta P$ is linear in $\nabla\cdot{\bf u}$, allowing the determination of the bulk viscosity using $\delta P=-B\nabla\cdot{\bf u}$.

The bulk viscosity was calculated assuming that the width $\Gamma=400$ MeV, and that the chemical equilibration time scales inversely with the density and $\tau_{\rm chem}$=1 fm/$c$ for a density of one quark per fm$^3$. 

\begin{figure}
\centerline{\includegraphics[width=8cm]{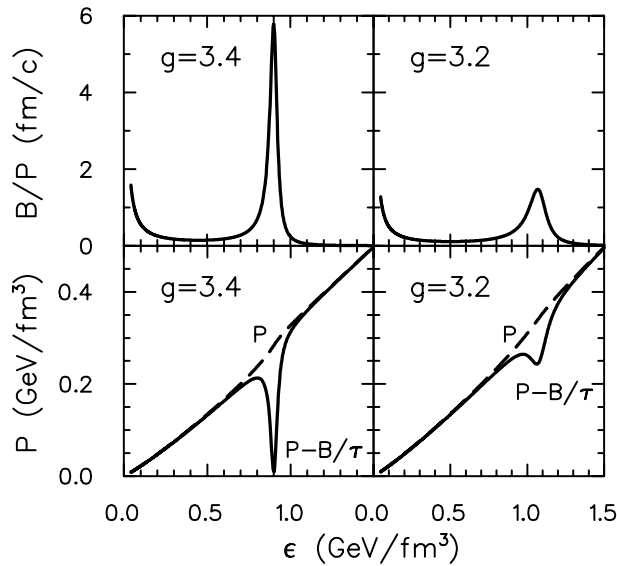}}
\caption{\label{fig:visctau}The bulk viscosity, scaled by the energy density, is  displayed in the upper panels for the linear sigma model. The peak at $T_c$ is due to the $\sigma$ field's inability to reach equilibrium, while the peak at low energy density is due to falling away from equilibrium. Viscous terms are larger and sharper for couplings close to the critical coupling ($g_c=3.554$). Lower Panel: For a Bjorken expansion ($\nabla\cdot{\bf u}=1/\tau$) the pressure is plotted alongside the Navier-Stokes expression, $T_{ii}=P-B\nabla\cdot{\bf u}$. Since the Navier-Stokes expression is only valid for small changes to the stress-energy tensor, the dynamics of the mean field should be handled explicitly if the corrections are large as are illustrated in the lower left-hand panel.}
\end{figure}
For a Bjorken expansion $\nabla\cdot{\bf u}=1/\tau$, and assuming an isentropic expansion starting with $\epsilon=8$ GeV/fm$^3$ at $\tau=1$ fm/$c$, we calculated both $P$ and $B$ as a function of $\tau$. To illustrate the size of the effect, we display both $P$ and the Navier-Stokes expression $T_{ii}=P-B\nabla\cdot{\bf u}$ as a function of the energy density for a Bjorken expansion in Fig. \ref{fig:visctau}. The effect is certainly dramatic for $g\approx g_c$, but since the Kubo relations are based on linear response theory, the large effect invalidates the underlying assumptions of Navier-Stokes hydrodynamics. We expect Israel-Stewart \cite{Israel:1979wp,Muronga:2004sf,Muronga:2003ta,Muronga:2001zk,Heinz:2005bw} equations for hydrodynamics  to result in moderated effects compared to Navier-Stokes, though they should give identical results if the corrections are modest. If the effects are also large in Israel-Stewart solutions, one should consider treating the dynamics of the mean field explicitly along the lines of \cite{Paech:2005cx}, where the equations of motion for hydrodynamics and for the the mean field were solved in parallel. 

\section{Summary}

The simplicity of the Kubo relations, Eq.s (\ref{eq:kuboshear}) and Eq. (\ref{eq:kuboB}), masks the wide variety of physical sources of viscosity. The one common aspect of the various sources is that non-zero equilibration times or non-zero interaction ranges can always be identified. We have focused on bulk viscosities associated with the chiral transition. In general, one would expect such effects whenever a system needs to rapidly rearrange its basic structure. In this sense these effects have much in common with super-cooling or hysteresis. In the case of a first order transition where super-cooling occurs, the departure from equilibrium is not proportional to the rate at which the system is changing, and the language of Navier-Stokes hydrodynamics is inappropriate.

The peaking of the bulk viscosity near $T_c$ is in stark contrast to the behavior of the shear viscosity for many fluids which comes to a minimum near $T_c$ \cite{Csernai:2006zz}. In \cite{Csernai:2006zz} convincing physical arguments are presented that the shear viscosity for the deconfinement transition also comes to a minimum near $T_c$. If the source of the viscosity is mainly due to the system's failure to equilibrate a scalar quantity such as the $\sigma$ field, one physically expects the singularity to be confined to the bulk viscosity. However, it is of interest that models of binary fluids also make predictions of a singularity in the shear viscosity near $T_c$ as described in \cite{hohenberghalperin}, where physical arguments are made by thermodynamically linking the diverging correlation length to a divergence in viscous forces. One lesson from the study of critical phenomena is that critical exponents inferred from mean-field models like those discussed here will likely be incorrect, even though the qualitative behavior might be well reproduced.

The implications for dynamics should be that the matter accelerates more quickly due to the higher gradients in $T_{xx}$ when the interior energy density is above the critical region. Once the matter flows into the viscous region  of energy densities, there should be a slowing down and a reduction of surface emission. This trend would be in the right direction to explain HBT measurements which show a rapid expansion with a sudden disintegration \cite{Retiere:2003kf}, but the potential magnitude of the effects are not yet known.

Finally, we re-emphasize that if one were to solve for the evolution of the mean fields or chemistry alongside solving the hydrodynamic evolution equations, one could forego incorporating these effects through the viscosity coefficients. If  the stress-energy tensor is strongly affected, the proper conclusion may be that rather than absorbing these effects into viscous hydrodynamics, one should treat non-equilibrated degrees of freedom explicitly. For instance, the dynamics of the $\sigma$ field can be calculated in parallel to the hydrodynamic equations of motion as was done in \cite{Paech:2005cx}, though it would be important to incorporate damping effects which were neglected there. Furthermore, chemical non-equilibration might be accounted for by solving for the time evolution of chemical abundances, then expressing the pressure in terms of the resulting non-equilibrated densities.

\appendix\section{Equivalence between Damped Harmonic Oscillator and Mean Field Equations}
\label{appendix}

The differential equations for the harmonic oscillator and for the Klein-Gordon equations,
\begin{eqnarray}
\label{eq:diffyq}
m\frac{d^2x}{dt^2}&=&-\gamma\frac{dx}{dt}-k(x-x_0(t))+F(t),\\
\nonumber
\frac{d^2\sigma}{dt^2}&=&-\Gamma\frac{d\sigma}{dt}-m_\sigma^2(\sigma-\sigma_{\rm eq})+F(t),
\end{eqnarray}
are equivalent after the substitutions $x\leftrightarrow\sigma$, $m\leftrightarrow 1$, $\gamma\leftrightarrow\Gamma$ and $k\leftrightarrow m_\sigma^2$. Here, $F(t)$ is an external random driving force, i.e., a Langevin force. After solving the more physically intuitive harmonic oscillator, a simple substitution will provide the answer for the Klein-Gordon equation.

Here, we consider the harmonic oscillator where the minimum of the potential moves with a velocity $v_0$, $x_0=v_0t$. Defining $x'\equiv x-x_0(t)$, one can find a differential equation,
\begin{equation}
m\frac{\partial^2x'}{\partial t^2}=\gamma\frac{\partial x'}{\partial t}
-kx'+F(t)-\gamma v_0.
\end{equation}
Thus, the effect of moving the potential is a drag force $\gamma v_0$, which causes the mean position to be offset from the center of the potential by an amount,
\begin{equation}
\label{eq:harmoscresult}
\delta x'=-\gamma v_0/k.
\end{equation}
This illustrates the importance of the drag term, and the irrelevance of the mass term or the strength of the random term.

The same result, Eq. (\ref{eq:harmoscresult}) can be derived from the classical expression for linear response theory,
\begin{eqnarray}
\label{eq:classicallinresponse}
\langle \delta x\rangle&=&\beta\int dt'\langle x(t=0)\partial_{t'}V(t')\rangle,\\
\nonumber
V(t)&=&\frac{1}{2}k[x-x_0(t)]^2-\frac{1}{2}kx^2\approx -kxv_0 t,
\end{eqnarray}
which leads to
\begin{eqnarray}
\label{eq:delxint}
\langle\delta x\rangle&=&(kv_0/T) \int_{-\infty}^{0} t' dt'
\langle x(0) \partial_{t'}x(t')\rangle\\
\nonumber
&=&-(kv_0/T) \int_{-\infty}^{0} dt'
\langle x(0) x(t')\rangle.
\end{eqnarray}
The solutions can be written in terms of the external force in frequency space using Green's functions,
\begin{eqnarray}
x(\omega)&=&\phi(\omega)F(\omega),\\
\phi(\omega)&=&\frac{1}{-m\omega^2-i\gamma\omega+k}.
\end{eqnarray}
The correlations in time can be written as:
\begin{equation}
\langle x(0)x(t)\rangle =
\int dt'dt'' \phi(-t')\phi(t-t'') \langle F(t') F(t'') \rangle,
\end{equation}
where for a Langevin force, $\langle F(t)F(t')\rangle =L\delta(t-t')$,
\begin{eqnarray} 
\langle x(0)x(t)\rangle &=& L \int dt' \phi(-t')\phi(t-t')\\
&=&\frac{L}{2\pi}\int d\omega e^{-i\omega t} \phi(\omega)\phi(-\omega).
\end{eqnarray}
The integration over time then is then
\begin{equation}
\int_{-\infty}^0 dt \langle x(0)x(t)\rangle =
L|\phi(\omega=0)|^2=\frac{L}{2k^2}.
\end{equation}
The strength of the Langevin force $L$ can be determined by calculating $\langle x(0)^2\rangle$, then using the equipartition theorem.
\begin{eqnarray}
\langle x^2 \rangle &=& L \int dt'[\phi(-t')]^2
=\frac{L}{2\pi}\int d\omega |\phi(\omega)|^2\\
\nonumber
&=&\frac{L}{2\pi m^2}\int d\omega \left\{
\frac{1}{(\omega-i\gamma/2m-\Omega)}
+\frac{1}{(\omega-i\gamma/2m+\Omega)}\right . \\
\nonumber &&\hspace*{60pt}\left .
+\frac{1}{(\omega+i\gamma/2m-\Omega)}
+\frac{1}{(\omega+i\gamma/2m+\Omega)}\right\},
~~\Omega=\sqrt{(k/m)^2+(\gamma/2m)^2}\\
\nonumber
&=& \frac{L}{2k\gamma},
\end{eqnarray}
which when combined with the equipartion theorem, $(1/2)k\langle x^2\rangle
=(1/2)T$, gives
\begin{equation}
L=2\gamma T.
\end{equation}
As expected, the strength of the random force depends on the temperature and the damping, and is independent of the mass or spring constant. When combined with Eq. (\ref{eq:classicallinresponse}) yields
\begin{equation}
\langle\delta x\rangle = -\frac{\gamma v_0}{k}
\end{equation}
which agrees with the simple statement above that on average the position of the mass lags behind the minimum of the potential by an amount such that the drag force, $\gamma v_0$, cancels the restoring force of the spring, $k\delta x$. It is remarkable that the result is independent of either the mass or the temperature. 

The zero mass limit can be linked to the diffusion equation. To make this connection, we consider a random force $F$ which acts for a time $\delta t$. In the zero-mass limit of Eq. (\ref{eq:diffyq}) particles move by an amount,
\begin{equation}
\delta x=-\left(\frac{k}{\gamma}x+\frac{F}{\gamma}\right)\delta t,
\end{equation}
which means the distribution changes due to the underlying translation,
\begin{eqnarray}
\delta f(x)&=&-f(x)\frac{\partial}{\partial x}\delta x
-\delta x\frac{\partial f(x)}{\partial x}
+\frac{1}{2}\delta x^2\frac{\partial^2 f(x)}{\partial x^2}\\
\nonumber
&=&\frac{k}{\gamma}\delta t f(x)
+\frac{kx}{\gamma}\delta t\frac{\partial f(x)}{\partial x}
+\frac{1}{2}\frac{(F^2+k^2x^2) \delta t^2}{\gamma^2}
\frac{\partial^2 f(x)}{\partial x^2},
\end{eqnarray}
where terms linear in $F$ have been discarded assuming that the sign of the force is random. Taking the limit $\delta t\rightarrow 0$, and assuming the Force is of the Langevin form, $F^2\delta t=L$, one finds,
\begin{eqnarray}
\frac{\partial f(x)}{\partial t}&=&\frac{k}{\gamma}f(x)+\frac{kx}{\gamma}
\frac{\partial f(x)}{\partial x}
+D\frac{\partial^2 f(x)}{\partial x^2},\\
\nonumber
D&=&\frac{L}{2\gamma^2}=\frac{T}{\gamma}.
\end{eqnarray}

Thus, if the dynamics of a field are described by either a first-order or second-order differential equation, the problem can be mapped to the harmonic oscillator. Furthermore, if it is first order as is the case for chemical equilibration, it can be mapped to the diffusion equation.

After making the substitutions described in Eq. (\ref{eq:diffyq}), the result translates to:
\begin{equation}
\langle\delta\sigma\rangle=\frac{m_\sigma^2}{\Gamma}\frac{d\sigma_{\rm eq}}{dt}.
\end{equation}
The similarity between the equations of motion for a harmonic oscillator and for fields allow for an easier understanding of the terms in the Klein-Gordon equation. The field's effective mass, $m_\sigma^2$, plays the role of the spring constant and provides the restoring force pushing the field towards equilibrium, whereas the width $\Gamma$ provides an effective drag force which impedes the field from maintaining equilibrium. This might be opposite to one's intuitive expectation: that a large width $\Gamma$ would indicate a quick decay of any excitation of the field.

\section*{Acknowledgments}
Support was provided by the U.S. Department of Energy, Grant No. DE-FG02-03ER41259. K.P. gratefully acknowledges the support of the Feodor Lynen program of the Alexander von Humboldt Foundation. The authors also wish to note inspiring discussions with Wolfgang Bauer and Misha Stephanov.


\begin{thebibliography}{99}
\bibitem{RHIC_Whitepapers}
  I.~Arsene {\em et al.},
  Nucl.\ Phys.\ A {\bf 757}, 1 (2005);
  %
  B.~B.~Back {\em et al.},
  {\em ibid.} {\bf 757}, 28 (2005);
  %
  J.~Adams {\em et al.},
  {\em ibid.} {\bf 757}, 102 (2005);
  %
  K.~Adcox {\em et al.},
  {\em ibid.} {\bf 757}, 184 (2005);

\bibitem{foerster} D. Forster, {\it Hydrodynamic Fluctuations, Broken Symmetry, and Corelation Functions}, Benjamin Cummings (1975).

\bibitem{Muronga:2003tb}
  A.~Muronga,
  Phys.\ Rev.\ C {\bf 69}, 044901 (2004)
  [arXiv:nucl-th/0309056].

\bibitem{Gelman:2006xw}
  B.~A.~Gelman, E.~V.~Shuryak and I.~Zahed,
  arXiv:nucl-th/0601029.

\bibitem{weinberg} S. Weinberg, {\it Gravitation and Cosmology}, John Wiley and Sons Inc. (1972).

\bibitem{Gyulassy:1997ib}
  M.~Gyulassy, Y.~Pang and B.~Zhang,
  Nucl.\ Phys.\ A {\bf 626}, 999 (1997)
  [arXiv:nucl-th/9709025].

\bibitem{Teaney:2003kp}
  D.~Teaney,
  Phys.\ Rev.\ C {\bf 68}, 034913 (2003)
  [arXiv:nucl-th/0301099].

\bibitem{Cheng:2001dz}
  S.~Cheng {\it et al.},
  Phys.\ Rev.\ C {\bf 65}, 024901 (2002)
  [arXiv:nucl-th/0107001].

\bibitem{Kortemeyer:1995di}
  G.~Kortemeyer, W.~Bauer, K.~Haglin, J.~Murray and S.~Pratt,
  Phys.\ Rev.\ C {\bf 52}, 2714 (1995)
  [arXiv:nucl-th/9509013].

\bibitem{Krasnitz:2002mn}
  A.~Krasnitz, Y.~Nara and R.~Venugopalan,
  Nucl.\ Phys.\ A {\bf 717}, 268 (2003)
  [arXiv:hep-ph/0209269].
    
\bibitem{Paech:2005cx}
  K.~Paech and A.~Dumitru,
  Phys.\ Lett.\ B {\bf 623}, 200 (2005)
  [arXiv:nucl-th/0504003].

\bibitem{Paech:2003fe}
  K.~Paech, H.~St\"ocker and A.~Dumitru,
  Phys.\ Rev.\ C {\bf 68}, 044907 (2003)
  [arXiv:nucl-th/0302013].

\bibitem{Israel:1979wp}
  W.~Israel, Annals Phys.\  {\bf 100}, 310 (1976);
  W.~Israel and J.~M.~Stewart,
  Annals Phys.\  {\bf 118}, 341 (1979).

\bibitem{Muronga:2004sf}
  A.~Muronga and D.~H.~Rischke,
  arXiv:nucl-th/0407114.

\bibitem{Muronga:2003ta}
  A.~Muronga,
  Phys.\ Rev.\ C {\bf 69}, 034903 (2004)
  [arXiv:nucl-th/0309055].

\bibitem{Muronga:2001zk}
  A.~Muronga,
  Phys.\ Rev.\ Lett.\  {\bf 88}, 062302 (2002)
  [Erratum-ibid.\  {\bf 89}, 159901 (2002)]
  [arXiv:nucl-th/0104064].

\bibitem{Heinz:2005bw}
  U.~Heinz, H.~Song and A.~K.~Chaudhuri,
  arXiv:nucl-th/0510014, Phys. Rev. C, in press.
  arXiv:nucl-th/0510038.

\bibitem{Csernai:2006zz}
  L.~P.~Csernai, J.~I.~Kapusta and L.~D.~McLerran,
  arXiv:nucl-th/0604032.

\bibitem{hohenberghalperin}
P.C. Hohenberg and B.I. Halperin, Rev. Mod. Phys., {\bf 49}, No. 3 (1977).

\bibitem{Retiere:2003kf}
  F. Retiere and M.A. Lisa, 
  Phys. Rev. C70, 044907 (2004).
\end{thebibliography}
\end{document}